\documentclass[sigconf, nonacm]{acmart}

\newcommand{\design}{TorR}

\usepackage{hyperref}
\usepackage{amsmath,amsfonts}
\usepackage{algorithm,algorithmic}
\usepackage{listings}
\usepackage{array}
\usepackage{graphicx}
\usepackage{textcomp}
\usepackage{xcolor}
\usepackage{orcidlink}
\usepackage{svg}
\usepackage{enumitem}
    
\usepackage{multirow}
\usepackage{makecell}
\usepackage{booktabs}

\newcolumntype{P}[1]{>{\centering\arraybackslash}p{#1}}
\newcolumntype{M}[1]{>{\centering\arraybackslash}m{#1}}

\usepackage{subcaption}
\usepackage{verbatim}
\newcommand\resetstackedplots{
\makeatletter
\pgfplots@stacked@isfirstplottrue
\makeatother
\addplot [forget plot,draw=none] coordinates{(1,0) (2,0) (3,0)};
}

\usepackage{tikz,pgfplots}
\usetikzlibrary{arrows.meta,positioning,fit}
\pgfplotsset{compat=1.18}
\usepackage{enumitem}
\usetikzlibrary{arrows.meta,positioning}

\begin{document}

\title{\design: Towards Brain-Inspired Task-Oriented Reasoning via Cache-Oriented Algorithm-Architecture Co-design}
\author{Hyunwoo Oh, SungHeon Jeong, Suyeon Jang, Hanning Chen,\\Sanggeon Yun, Tamoghno Das and Mohsen Imani}
\affiliation{\institution{Department of Computer Science, University of California, Irvine} \city{Irvine}\state{CA}\country{USA}}
\email{{hyunwooo, m.imani}@uci.edu}


\begin{abstract}
Task-oriented object detection (TOOD) atop CLIP offers open-vocabulary, prompt-driven semantics, yet dense per-window computation and heavy memory traffic hinder real-time, power-limited edge deployment. We present \emph{\design{}}, a brain-inspired \textbf{algorithm--architecture co-design} that \textbf{replaces CLIP-style dense alignment with a hyperdimensional (HDC) associative reasoner} and turns temporal coherence into reuse. On the \emph{algorithm} side, \design{} reformulates alignment as HDC similarity and graph composition, introducing \emph{partial-similarity reuse} via (i) query caching with per-class score accumulation, (ii) exact $\delta$-updates when only a small set of hypervector bits change, and (iii) similarity/load-gated bypass under high system load. On the \emph{architecture} side, \design{} instantiates a lane-scalable, bit-sliced item memory with bank/precision gating and a lightweight controller that schedules bypass/$\delta$/full paths to meet RT-30/RT-60 targets as object counts vary. Synthesized in a TSMC 28\,nm process and exercised with a cycle-accurate simulator, \design{} sustains real-time throughput with millijoule-scale energy per window ($\approx$50\,mJ at 60\,FPS; $\approx$113\,mJ at 30\,FPS) and low latency jitter, while delivering competitive AP@0.5 across five task prompts (mean 44.27\%) within a bounded margin to strong VLM baselines, but at orders-of-magnitude lower energy. The design exposes deployment-time configurability (effective dimension $D'$, thresholds, precision) to trade accuracy, latency, and energy for edge budgets.
\end{abstract}
\maketitle

\vspace{-1mm}

\section{Introduction}
Many edge devices must make sense of the world while obeying tight latency and power limits: a home robot asked to “find something to scoop soup,” a wearable that helps a user “look for a cup,” or a drone that must “bring the object used to cut rope.” Vision–language models (VLMs) such as CLIP align images and text in a shared embedding space and enable this kind of open-vocabulary, promptable behavior \cite{clip}. Building on CLIP, task-oriented object detection (TOOD) reframes detection as selecting objects that fulfill a natural-language goal rather than a fixed label list \cite{taskclip,toist,cotdet,tood}.

\begin{figure}[t]
\centering
\includegraphics[width=\linewidth]{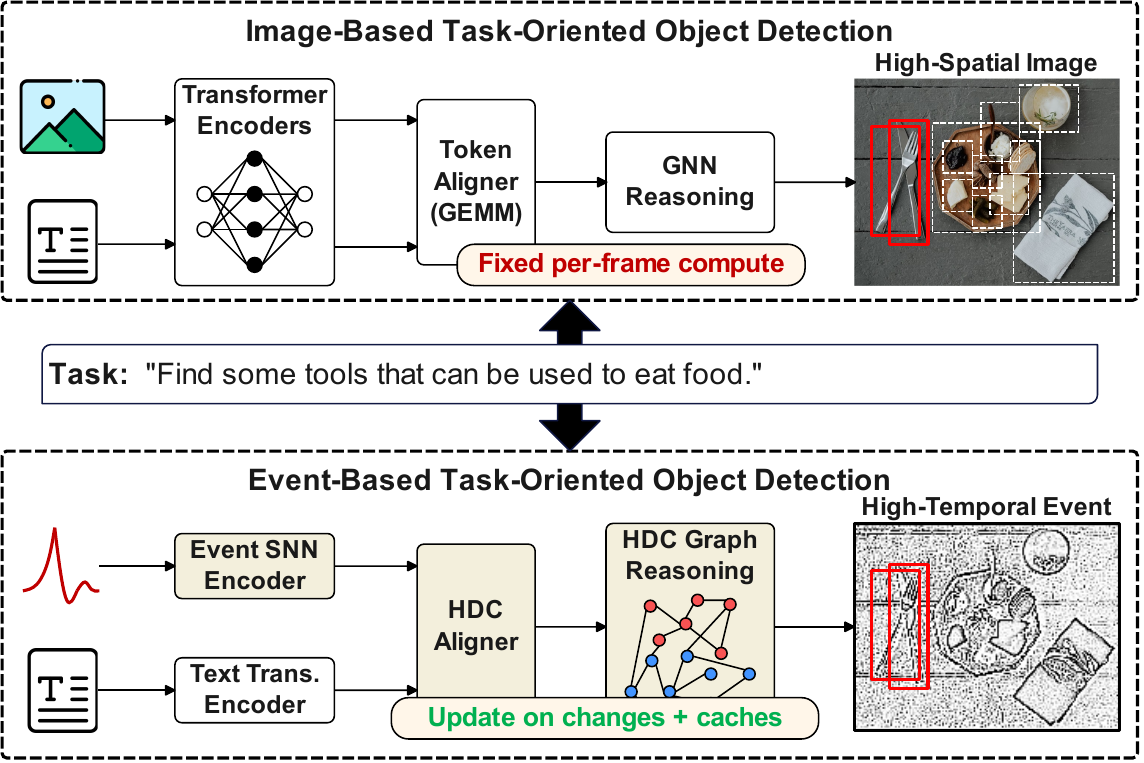}
\vspace{-5mm}
\caption{From CLIP/ViT to \design. \textbf{Top:} dense token aligner with fixed per-frame cost. \textbf{Bottom:} an event-driven encoder paired with an HDC associative aligner and a lightweight reasoner. Query caching turns temporal coherence into reuse so \design{} \emph{updates on change}.}
\label{fig:intro_overview}
\vspace{-5mm}
\end{figure}

Today’s pipelines typically inherit image-based Vision Transformers (ViTs) as their backbones. These attention/MLP stacks perform nearly the same amount of work every frame, independent of how much the scene changes \cite{vit}. As resolutions grow, activations spill off-chip; memory movement, not arithmetic, dominates latency and energy. Datacenters hide this cost with large batches and abundant bandwidth. Edge devices cannot: they process a live stream, one moment at a time, and their budgets are effectively frame-rate-locked. Worse, the pipeline usually treats each frame as a fresh problem, discarding what it learned a moment ago. Figure~\ref{fig:intro_overview} (top) sketches this status quo.

We take a different path inspired by how brains work: compute when the world changes, and reason by recalling what you already know. Instead of redoing the same dense work every frame, we produce embeddings that \emph{drift} gradually over time, then \emph{reuse} prior computation wherever possible. Event-driven encoders naturally support this “compute on change” view while hyperdimensional computing (HDC) provides a compact, noise-tolerant substrate for representing concepts and relations as ultra-high-dimensional hypervectors and matching them via simple associative similarity \cite{kanerva2009,rahimi2016}. These operations map cleanly to hardware and favor data reuse over raw FLOPs. Figure~\ref{fig:intro_overview} (bottom) illustrates our approach.

The core idea is to turn temporal coherence into efficiency. We cache what was inferred in the previous instant and, when the current scene is similar, update only what changed instead of recomputing from scratch. A lightweight controller decides whether to (i) aggressively reuse the cached result when similarity is very high and load is heavy, (ii) refresh scores by updating the portions that changed, or (iii) fall back to a full refresh when the scene really is new. An HDC graph reasoner then injects task knowledge—relations like \emph{used-for} or \emph{part-of}—so the system can prioritize objects that help achieve the user’s goal. The result is a pipeline whose cost follows the dynamics of the scene rather than the worst-case model size, and whose reasoning maps well to edge hardware.

This paper introduces \design, a cache-oriented algorithm–architecture co-design for task-oriented detection at the edge. \design{} pairs a lightweight, event-driven encoder with an HDC associative aligner and a lightweight reasoner. Instead of treating every frame as a reset, \design{} carries forward state, reuses partial results when the scene is stable, and expends effort only where evidence is new. To our knowledge, \design{} is the first end-to-end, brain-inspired pipeline for task-oriented detection explicitly co-designed around temporal reuse.

\noindent\textit{We contribute:}
\begin{itemize}[leftmargin=*,nosep]
  \item \textbf{Reuse-centric, similarity-gated pipeline.} We treat consecutive queries as incremental updates and use a simple policy to choose aggressive reuse, partial refresh, or full refresh—aligning work with scene dynamics and frames-per-second (FPS) budgets (30/60\,FPS).
  \item \textbf{Memory-friendly hyperdimensional substrate.} An HDC associative aligner and lightweight reasoner operate over on-chip caches and bit-sliced item memory, keeping compute close to data and minimizing off-chip traffic.
  \item \textbf{Experimental evidence.} On task-oriented detection, \design{} sustains 30/60\,FPS with strong AP@0.5, reducing alignment traffic and energy versus an SNN\,+\,naïve HDC baseline and maintaining accuracy under aggressive reuse.
\end{itemize}
\section{Background and Motivation}
\label{sec:background}

\subsection{Task-Oriented Detection on VLMs}
CLIP aligns images and text in a shared embedding space, enabling open-vocabulary behavior via image–text similarity \cite{clip}. Building on CLIP, \emph{task-oriented object detection (TOOD)} selects objects that \emph{fulfill a natural-language goal} rather than a fixed label list; recent systems include TaskCLIP, TOIST, and CoTDet \cite{taskclip,toist,cotdet,tood}. Most pipelines retain \emph{Vision Transformers (ViTs)} for perception \cite{vit}, whose attention/MLP/normalization impose largely \emph{fixed per-frame} compute and heavy activation traffic—fine for datacenters but ill-matched to edge latency/energy limits—motivating a bit-parallel, reuse-friendly alternative to dense CLIP-style alignment.

\subsection{Event-Driven Perception: DVS \& SNNs}
\emph{Dynamic Vision Sensors (DVS)} emit asynchronous events on log-intensity changes, offering microsecond latency and high temporal resolution at modest spatial resolution \cite{dvs_survey,lichtsteiner2008,davis2014}. Because there is no intrinsic frame rate, embedded systems aggregate events into windows of width \(\Delta t\) for scheduling \cite{dvs_survey}. \emph{Spiking neural networks (SNNs)} process such streams natively; neuromorphic platforms (e.g., Intel Loihi) achieve real-time, low-power inference by activating only where/when spikes occur \cite{loihi}. Replacing a ViT with an SNN thus removes dense, clocked vision compute and yields temporally coherent (drifting) embeddings that a reuse-first aligner can \emph{update} rather than recompute.

\subsection{Hyperdimensional Computing for Alignment/Reasoning}
\emph{Hyperdimensional computing (HDC)} represents symbols/structure as ultra-high-dimensional hypervectors; composition (binding/bundling) relies on simple, massively parallel operations, and retrieval reduces to similarity search \cite{kanerva2009,rahimi2016}. HDC arithmetic is bit-simple and noise-tolerant, and has even been executed in-memory \cite{karunaratne2020}. However, when used naïvely as a CLIP replacement, HDC becomes \emph{memory-bound}: each query scans many long hypervectors, so bandwidth—not arithmetic—sets runtime/energy; FPGA designs such as FACH mitigate this via restructuring and partial-result reuse \cite{imani2019fach}, pointing to \emph{data movement and reuse} as the true optimization target.

\begin{figure}[t]
\centering
\includegraphics[width=\linewidth]{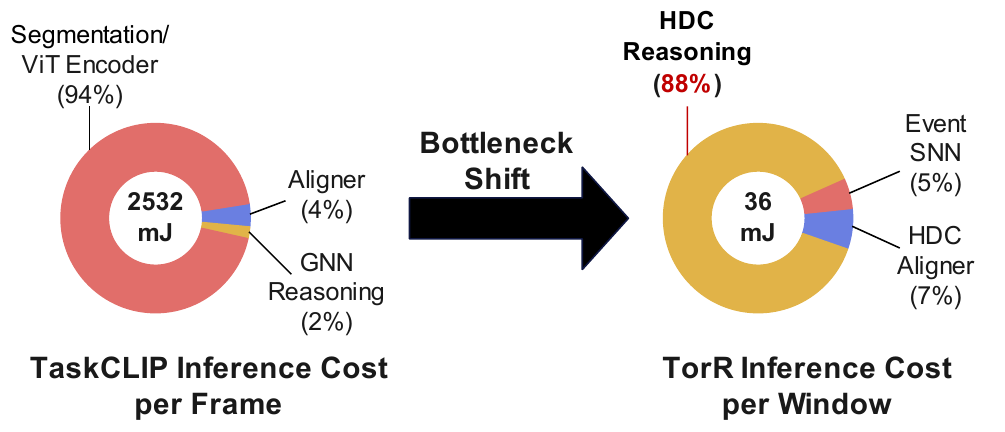}
\vspace{-4mm}
\caption{\textbf{Bottleneck shift from CLIP/ViT to \design.}
Left: TaskCLIP is dominated by the ViT backbone.
Right: after ViT\(\rightarrow\)event encoder, cost shifts to \emph{HDC associative search + graph reasoning}, which are memory-bound.}
\label{fig:workload_analysis}
\vspace{-3mm}
\end{figure}

\subsection{Motivation}
Once the ViT cost is removed (Fig.~\ref{fig:workload_analysis}), the \emph{associative search + reasoning} path dominates and is memory-bound, shifting the focus from FLOPs to \emph{data movement and reuse}. Adjacent event windows yield \emph{similar} queries; recomputing full scans wastes temporal coherence. We therefore co-design an \emph{HDC associative aligner} with cosine similarity and \(\delta\)-updates (partial similarity) plus a cache-oriented substrate—bit-sliced item memory and per-class accumulators—with deployment-time knobs (dimension \(D\), \(\delta\) budget, similarity thresholds, bank/precision gating), so cost tracks scene dynamics rather than worst-case model size.

\section{Algorithm-Architecture Co-Design}
\label{sec:codesign}

\subsection{Co-Design Overview}
\label{sec:overview}
\begin{figure}[t]
  \centering
  \includegraphics[width=0.87\linewidth]{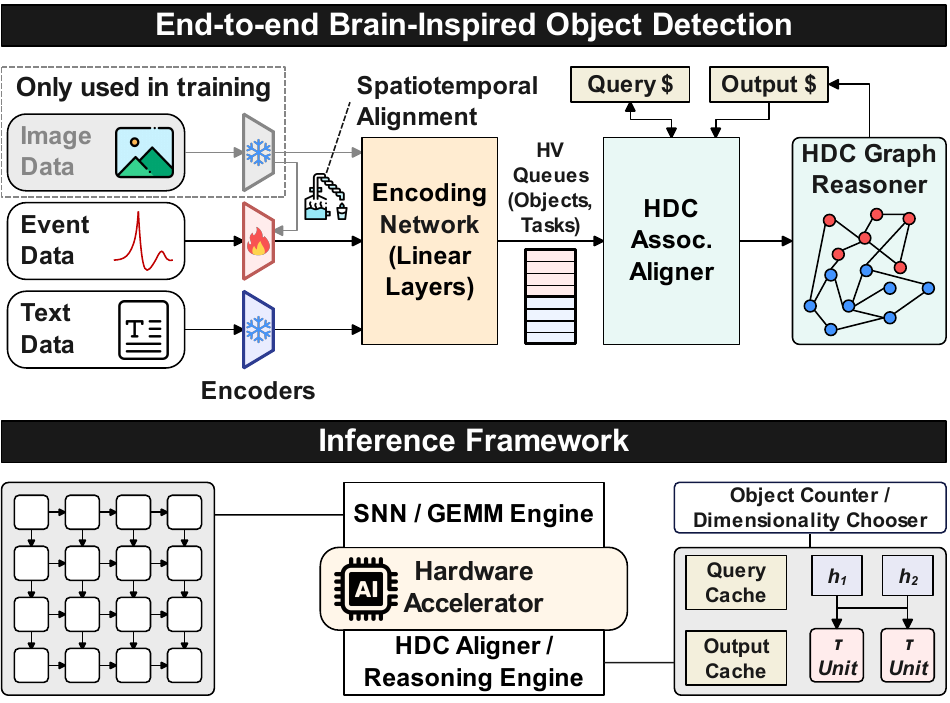}
  \vspace{-1mm}
  \caption{\textbf{\design{} overview.} Images are used only to transfer CLIP semantics to events. At run time, partial-similarity reuse (\(\delta\)-updates, caches) and FPS/QoS control align cost with scene dynamics.}
  \label{fig:codesign_overview}
  \vspace{-3mm}
\end{figure}

Figure~\ref{fig:codesign_overview} sketches \design. During \emph{training}, an event SNN is aligned to the image/text spaces so that event windows and RGB frames of the same object/task co-locate. At \emph{inference}, only the event SNN and text encoder are active. The SNN produces a query hypervector \(\mathbf{q}\); a similarity-gated controller (Alg.~\ref{alg:path-policy}) selects among \textsf{bypass}, \textsf{\(\delta\)-update}, and \textsf{full} paths. In \textsf{\(\delta\)-update}/\textsf{full}, an \emph{associative aligner} computes cosine scores \(s_j=\cos(\mathbf{q},\mathbf{h}_j)\) against an item-memory bank of concept HVs \(\{\mathbf{h}_j\}_{j=1}^{M}\); if not bypassed, an \emph{HDC graph reasoner} applies precomputed task weights \(\tilde w_j=\cos(\mathbf{g}_P,\mathbf{h}_j)\) to yield final scores \(\hat{s}_j=s_j\tilde w_j\). A query cache and \(\delta\)-updates exploit frame-to-frame similarity, while an FPS/QoS controller gates the effective dimension \(D'\) (bank gating) to meet 30/60\,FPS under dynamic loads.

\subsection{Algorithmic Design}
\label{sec:algo}

\textbf{Event SNN encoder.}
DVS events \(\mathcal{E}=\{(x,y,t,p)\}\) are aggregated over a window of width \(\Delta t\) to form a spatiotemporal tensor for a lightweight spiking backbone. Per proposal, the encoder outputs \(\mathbf{z}_e\!\in\!\mathbb{R}^d\), mapped to a bipolar HV by a fixed projection \(R\) and sign: \(\mathbf{q}=\mathrm{sign}(R\mathbf{z}_e)\in\{-1,+1\}^D\). The text prompt is encoded once and mapped to \(\mathbf{t}\) (precomputed under fixed tasks).

\begin{figure}[t]
\centering
\includegraphics[width=\linewidth]{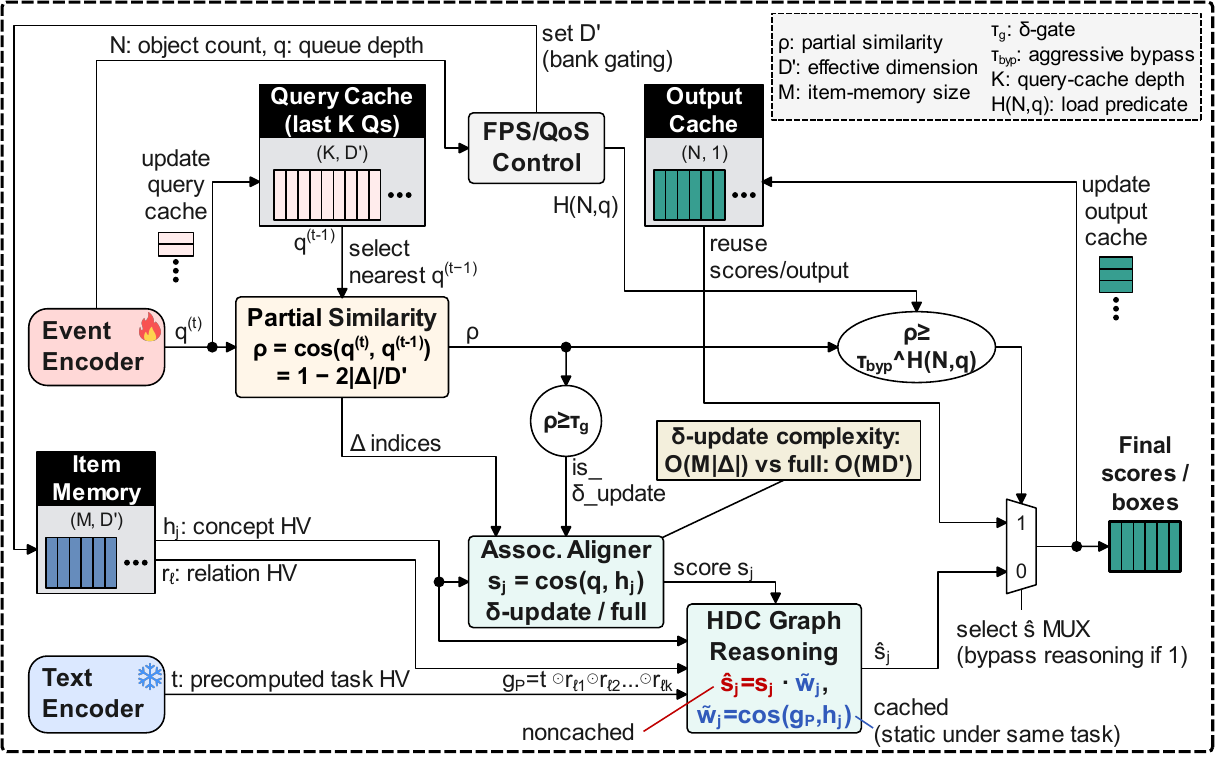}
\vspace{-4mm}
\caption{\textbf{Cache-gated HDC reasoner with \(\delta\)-alignment.}
A query cache (depth \(K\)) supplies the nearest prior query \(\mathbf{q}^{(t-1)}\). Partial similarity \(\rho=\cos(\mathbf{q}^{(t)},\mathbf{q}^{(t-1)})=1-\tfrac{2|\Delta|}{D'}\) selects \textsf{\(\delta\)-update} (update only flipped indices \(\Delta\)) or \textsf{full}. Under high load, if \(\rho\!\ge\!\tau_{\mathrm{byp}}\!\wedge\!H(N,q)\), cached scores/outputs are reused. Otherwise the aligner computes \(s_j=\cos(\mathbf{q},\mathbf{h}_j)\) and the reasoner (for fixed task) applies cached weights \(\tilde w_j=\cos(\mathbf{g}_P,\mathbf{h}_j)\) with \(\mathbf{g}_P=\mathbf{t}\odot r_{\ell_1}\odot\cdots\odot r_{\ell_k}\), producing \(\hat s_j=s_j\tilde w_j\). The FPS/QoS controller gates \(D'\) (bank gating).}
\label{fig:hdc_reasoner}
\vspace{-4mm}
\end{figure}

\textbf{Small training bridge (image\(\rightarrow\)event).}
We apply a light contrastive transfer so event features sit near image features in CLIP space while preserving text alignment. With frozen CLIP encoders \(f_{\text{img}},f_{\text{text}}\) and trainable SNN \(f_{\text{evt}}\), for frame \(I\) and window \(W(t_0){=}\{(t,p)\mid t_0{\le}t{<}t_0{+}\Delta t\}\):
\begin{equation}
\widetilde{E}(x,y)=\!\!\sum_{(t,p)\in W(t_0)}\!E(x,y,t,p),
\widehat{E}=\frac{\widetilde{E}}{\max_{x,y}|\widetilde{E}(x,y)|+\epsilon}
\label{eq:eventacc}
\end{equation}
With cosine \(s(\cdot,\cdot)\) and temperatures \(\tau_c,\tau_t\):
\begin{equation}
\mathcal{L}_{\text{con}}
= -\log \frac{\exp(s(f_{\text{img}}(I),f_{\text{evt}}(\widehat{E}))/\tau_c)}
{\sum_{e'\in\mathcal{B}}\exp(s(f_{\text{img}}(I),f_{\text{evt}}(e'))/\tau_c)},
\label{eq:infonce}
\end{equation}
\begin{equation}
\mathcal{L}_{\text{zs}}
= -\log \frac{\exp(s(f_{\text{evt}}(\widehat{E}),f_{\text{text}}(T))/\tau_t)}
{\sum_{k\in\mathcal{V}}\exp(s(f_{\text{evt}}(\widehat{E}),f_{\text{text}}(T_k))/\tau_t)},
\label{eq:zsl}
\end{equation}
and \(\mathcal{L}=\mathcal{L}_{\text{con}}+\alpha\,\mathcal{L}_{\text{zs}}\).

\textbf{Associative aligner.}
Item memory stores \(\{\mathbf{h}_j\}_{j=1}^{M}\subset\{-1,+1\}^{D}\).
With effective dimension \(D_{\mathrm{eff}}=D'\) (active banks), cosine scores are
\begin{equation}
s_j=\cos(\mathbf{q},\mathbf{h}_j)
=\frac{\langle\mathbf{q},\mathbf{h}_j\rangle}{\|\mathbf{q}\|\,\|\mathbf{h}_j\|}
=\frac{1}{D_{\mathrm{eff}}}\sum_{i=1}^{D_{\mathrm{eff}}}q_i h_{j,i},
\label{eq:sim}
\end{equation}
since \(\|\mathbf{q}\|=\|\mathbf{h}_j\|=\sqrt{D_{\mathrm{eff}}}\) for bipolar vectors.

\textbf{Query cache \& similarity gate (Fig.~\ref{fig:hdc_reasoner}).}
Among the last \(K\) queries, the cache supplies the nearest \(\mathbf{q}^{(t-1)}\).
Let \(\Delta=\{i:q^{(t)}_i\neq q^{(t-1)}_i\}\). Then
\begin{equation}
\rho=\cos(\mathbf{q}^{(t)},\mathbf{q}^{(t-1)})
=\frac{1}{D_{\mathrm{eff}}}\sum_{i=1}^{D_{\mathrm{eff}}}q^{(t)}_i q^{(t-1)}_i
=1-\frac{2|\Delta|}{D_{\mathrm{eff}}}.
\label{eq:gate}
\end{equation}
If \(\rho\ge\tau_g\) choose \textsf{\(\delta\)-update}; else perform \textsf{full} and refresh the cache. Alg.~\ref{alg:path-policy} combines this with load to select the path and \(D'\).

\textbf{Partial-similarity reuse (\(\delta\)-updates).}
Maintain per-class accumulators and update only flipped indices:
\begin{equation}
\forall j:\quad
s_j \leftarrow s_j + \frac{1}{D_{\mathrm{eff}}}\sum_{i\in\Delta}
\big(q^{(t)}_i - q^{(t-1)}_i\big)\,h_{j,i},
\label{eq:delta}
\end{equation}
where \(q^{(t)}_i - q^{(t-1)}_i\in\{\pm2\}\). Work drops from \(O(MD')\) to \(O(M|\Delta|)\).

\textbf{HDC graph reasoner.}
Relations \(\{r_\ell\}\) use Hadamard binding (\(\odot\)). A \(k\)-hop path
\(P=(\ell_1,\ldots,\ell_k)\) forms \(\mathbf{g}_P=\mathbf{t}\odot r_{\ell_1}\odot\cdots\odot r_{\ell_k}\).
For fixed tasks, reasoner weights are precomputed as \(\tilde w_j=\cos(\mathbf{g}_P,\mathbf{h}_j)\) and applied to aligner scores to yield \(\hat s_j=s_j\,\tilde w_j\).
A MUX after the reasoner selects between \textsf{bypass}/\textsf{aligner-only} and \textsf{aligner\(\times\)reasoner} outputs.

\textbf{Aggressive bypass.}
Under high load, if \(\rho \ge \tau_{\mathrm{byp}}\wedge H(N,q)\) the controller reuses cached scores/outputs and skips both \(\delta\)-update and the reasoner.

\textbf{FPS/QoS controller.}
Given object count \(N\), similarity \(\rho\), and queue depth \(q\), the controller implements Alg.~\ref{alg:path-policy}: it selects \textsf{bypass}/\textsf{\(\delta\)-update}/\textsf{full} and gates \(D'\) (bank gating) to respect the FPS budget.

\begin{algorithm}[h]
\small
\caption{Path policy (similarity-only controller)}
\label{alg:path-policy}
\begin{algorithmic}[1]
\REQUIRE similarity \(\rho\), object count \(N\), queue depth \(q\)
\REQUIRE thresholds \(\tau_{\mathrm{byp}}\), \(\tau_g\); load thresholds \(N_{\mathrm{hi}}\), \(q_{\mathrm{hi}}\)
\STATE \(H \leftarrow (N \ge N_{\mathrm{hi}}) \ \textbf{or}\ (q \ge q_{\mathrm{hi}})\) \hfill // high-load predicate
\IF{\(\rho \ge \tau_{\mathrm{byp}}\) \textbf{and} \(H\)}
  \STATE \textbf{return} \textsf{bypass} \hfill // reuse cached scores/outputs
\ELSIF{\(\rho \ge \tau_g\)}
  \STATE action \(\leftarrow\) \textsf{\(\delta\)-update}
\ELSE
  \STATE action \(\leftarrow\) \textsf{full}
\ENDIF
\STATE Choose effective dimension \(D'\) (bank gating) to meet FPS given \(N,q\)
\STATE \textbf{return} action, \(D'\)
\end{algorithmic}
\end{algorithm}

\section{Hardware Architecture}
\label{sec:hw}

The accelerator realizes the cache–gated hyperdimensional pipeline on a standard RTL substrate. Each window begins by comparing the current query hypervector \(\mathbf q^{(t)}\) with the nearest cached query \(\mathbf q^{(t-1)}\). The resulting similarity \(\rho\) and flipped-index set \(\Delta\) drive a small controller that selects the execution path and gates the effective hypervector dimension \(D'\). A lane-parallel associative aligner then computes cosine scores against the item memory; when evidence is stable, a cached output is returned, otherwise a lightweight reasoner applies task weights and commits results to the output cache.

\subsection{Top-level overview}
\begin{figure}[t]
  \centering
  \includegraphics[width=\linewidth]{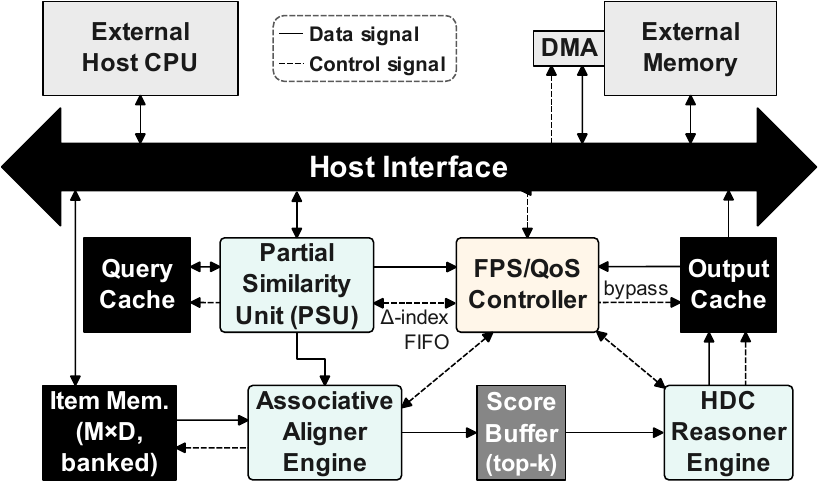}
  \vspace{-3mm}
  \caption{\textbf{Similarity-gated top-level accelerator.}
  PSU computes inter-query similarity \(\rho\). The FPS/QoS controller selects bypass/\(\delta\)/full, gates \(D'\) (bank enables) and precision, and programs the associative aligner. In \(\delta\)-mode the aligner uses a \(\Delta\)-index FIFO for sparse reads from the banked item memory \(M\times D\); scores are accumulated, top-\(k\) pooled, optionally reasoned (HDC), cached, and returned via host/DMA.
  Solid arrows = data, dashed = control.}
  \label{fig:top_arch}
  \vspace{-3mm}
\end{figure}

Figure~\ref{fig:top_arch} shows the datapath and control. A query cache stores the last \(K\) queries; the partial-similarity unit (PSU) produces \(\rho\) and fills a \(\Delta\)-index FIFO with flipped positions. The FPS/QoS controller uses \(\rho\) and load (object count and queue depth) to choose among bypass, delta, and full execution. It also enables a subset of bit-sliced banks in the item memory to realize \(D'\), selects accumulator precision, and sets the normalization shift \(\log_2 D'\). The associative aligner reads concept hypervectors from the enabled banks and produces a score vector. A top-\(k\) key and margin gate reasoning: when both match the previous window, reasoning is skipped and the cached output is forwarded; otherwise the reasoner multiplies scores by precomputed task weights and updates the output cache before DMA to the host.

\subsection{Shared similarity micro-kernel}
All engines reuse the same micro-kernel for bipolar cosine. Bitwise XNOR implements \(\pm 1\) multiplication, a short adder tree popcounts matches versus mismatches, and a fixed right shift by \(\log_2 D'\) applies normalization. The kernel supports two access patterns that map directly to the algorithm: streaming reads for full scans and sparse reads indexed by \(\Delta\) for delta updates. Per-class accumulators persist across windows so sparse corrections apply without recomputing unchanged columns. Bank enables are honored on every read, so \(D'\) acts as a runtime QoS knob.

\subsection{Associative cosine aligner}
The aligner computes
\[
s_j=\frac{1}{D'}\sum_{i=1}^{D'} q_i h_{j,i},\quad
\mathbf q,\mathbf h_j\in\{-1,+1\}^{D'}.
\]
In full mode the index generator streams one column per cycle from the enabled banks; each column is broadcast to \(W\) class lanes that perform XNOR\(\rightarrow\)popcount and accumulate into on-chip registers. In delta mode the aligner pops indices from the \(\Delta\)-FIFO and touches only those columns. Because accumulators persist across windows, a flipped bit contributes a signed correction of magnitude \(2/D'\) in cosine space, implemented as a \(\pm 2\) update in the integer domain followed by the final normalization shift. With \(B\) banks and \(W\) lanes, the latency scales as
\[
\text{cycles}_{\text{full}} \approx D'\,\Big\lceil\frac{M}{W}\Big\rceil,
\qquad
\text{cycles}_{\delta} \approx |\Delta|\,\Big\lceil\frac{M}{W}\Big\rceil,
\]
and memory traffic drops from \(O(MD')\) to \(O(M|\Delta|)\) when queries change little.

\subsection{Partial-similarity unit}
The PSU detects query drift and supplies sparse indices. XOR against the cached query identifies flipped bits; a popcount yields \(|\Delta|\) and an affine map produces \(\rho=1-2|\Delta|/D'\). The PSU writes \(\Delta\) to the index FIFO for the aligner’s delta mode and forwards \(\rho\) to the controller. This converts temporal coherence into concrete savings by steering the aligner toward sparse updates and, under high load with high similarity, toward bypass.

\subsection{Reasoner and cache gating}
The reasoner scales \(s_j\) by precomputed task weights \(w_j\) stored on chip. For fixed prompts these weights are computed once offline; at run time the reasoner reduces to a vector MAC with rounding and saturation. When the top-\(k\) key and margin match the previous window, the reasoner remains gated and the cached output is reused. If prompts change online, the same similarity kernel can recompute \(w_j\) by treating the prompt hypervector as a query.

\subsection{Policy to hardware controls}
The controller maps Algorithm~\ref{alg:path-policy} to a small set of window-level registers: mode selects bypass, delta, or full; bank-enable bits define \(D'\); a precision bit selects int8 or int4 accumulators; the normalization shift equals \(\log_2 D'\); head and tail pointers drive the \(\Delta\)-index FIFO. Controls are latched once per window, keeping the datapath feed-forward and timing-robust. The same register file configures the PSU and reasoner enables so that alignment always precedes any decision to reuse reasoning.

\subsection{Bandwidth, energy, and timing}
Full scans read \(M D'\) bits per window from the item memory, while delta reads \(M|\Delta|\) bits plus \(|\Delta|\) indices. With banking, each bank serves about \(M|\Delta|/B\) bits in delta mode. Clock and bank gating follow the selected mode and bank mask, so dynamic power scales with \(|\Delta|\) and \(D'\). The aligner and reasoner are fully pipelined to produce one column per cycle in full mode, one flipped column per cycle in delta mode, and one score product per lane per cycle in the reasoner. Cosine normalization is a shift, so no divider appears on the critical path.

\subsection{Interfaces and state}
The host interface ingests queries and returns scores and boxes via DMA. Window-level state consists of the query cache, the \(\Delta\)-index FIFO, per-class accumulators, and the output cache. A small FSM sequences: latch controls, run aligner in the selected mode, optionally run the reasoner, update caches, and emit results. This arrangement keeps the top-level simple while concentrating complexity inside the shared similarity kernel.

\section{Evaluation}
\label{sec:eval}

\subsection{Experimental Setup}
We evaluate five task-oriented prompts representative of everyday activities:
\emph{1. pour wine into a glass}, \emph{2. sports}, \emph{3. cooking}, \emph{4. have breakfast}, and \emph{5. take a rest}.
Accuracy is measured as AP@0.5 (IoU=0.5). System metrics are end-to-end latency, throughput (FPS), power, and energy per frame.
All execution results are produced by a cycle-accurate simulator that replays our workloads and is calibrated to switching activity and timing
from ASIC synthesis of our accelerator written in Verilog HDL using TSMC 28\,nm at 1\,GHz. The accelerator offers run-time QoS via bank-gated effective dimension ($D'$), and a cache-aided partial-update path governed by a lightweight controller that enforces 30/60 FPS targets, defined as RT-30/RT-60.

\begin{figure}[t]
\centering
\includegraphics[width=1\linewidth]{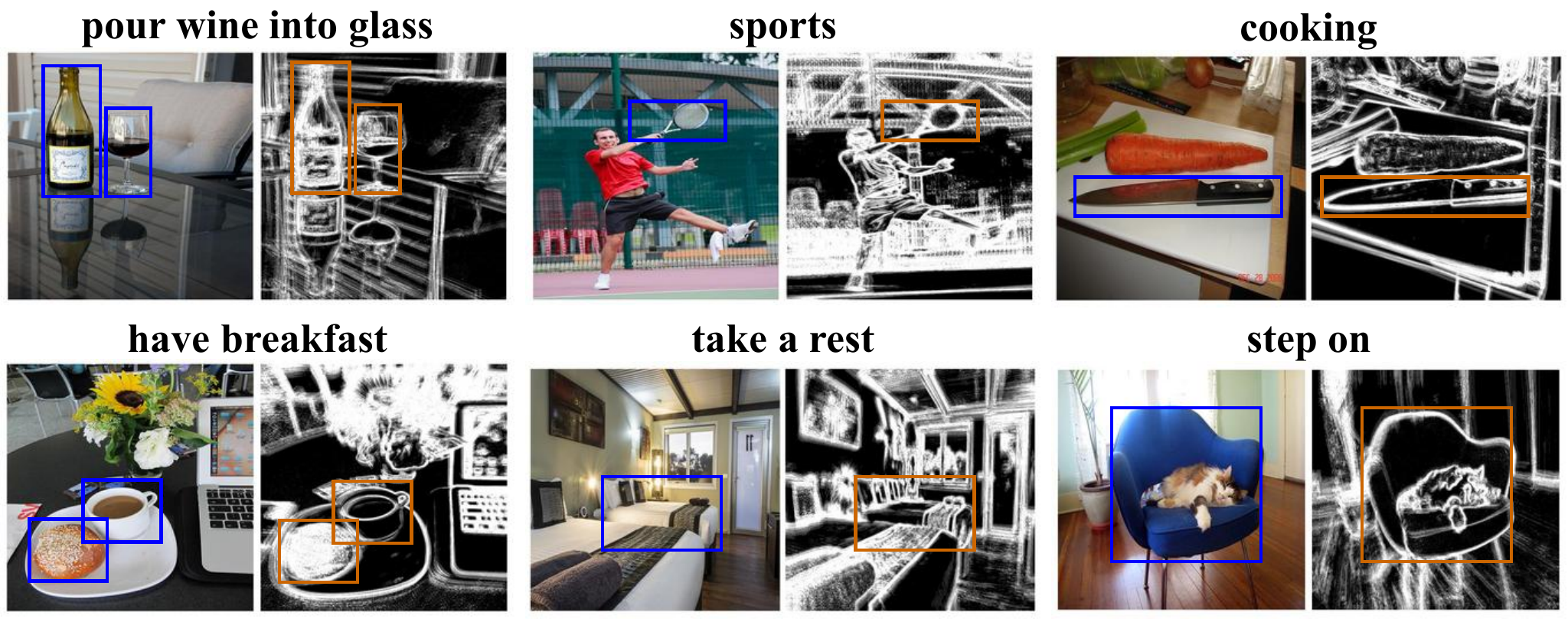}
\vspace{-6mm}
\caption{\textbf{Five example tasks for TOOD evaluation.} Detected objects are highlighted in blue.}
\label{fig:eval_tasks}
\vspace{-4mm}
\end{figure}

\subsection{ASIC Synthesis (28\,nm)}
We synthesize the Verilog RTL with Synopsys Design Compiler in a topographical flow (TT, 1.0\,V, 1\,GHz). Logic power is taken from the synthesis power model and later exercised with the cycle-accurate activity traces used in our execution experiments; SRAM figures come from compiled 28\,nm views. The design is intentionally compute-centric: most silicon is spent in the associative aligner where similarity is evaluated, while control, sorting, and I/O remain lightweight. Table~\ref{tab:hw_spec_combined} summarizes the full hardware footprint without low-level operator bookkeeping.

\begin{table}[t]
  \centering
  \small
  \caption{Hardware specifications (TSMC 28\,nm, 1\,GHz). Logic areas in mm$^2$, powers in mW.}
  \vspace{-2mm}
  \label{tab:hw_spec_combined}
  \begin{tabular}{lcc}
    \toprule
    \textbf{Block / Component} & \textbf{Area (mm$^2$)} & \textbf{Power (mW)} \\
    \midrule
    \multicolumn{3}{l}{\textit{Logic (synthesized)}} \\
    Associative Aligner            & 4.488 & 3{,}522.56 \\
    Lightweight Reasoner           & 0.642 & 504.32 \\
    Partial-Update Unit            & 0.280 & 220.16 \\
    Score Buffer (top-$k$)         & 0.140 & 110.08 \\
    Sorter                         & 0.140 & 110.08 \\
    Controller (RT/QoS)            & 0.070 & 55.04 \\
    Host IF / DMA                  & 0.105 & 82.56 \\
    $\Delta$-index FIFO \& misc.   & 0.070 & 55.04 \\
    \textbf{Total (logic)}         & \textbf{5.937} & \textbf{4{,}659.84} \\
    \midrule
    \multicolumn{3}{l}{\textit{SRAM macros (compiled)}} \\
    Item memory (banked)           & 0.50  & 120 \\
    Query/Output caches            & 0.03  & 15 \\
    \textbf{Total (SRAM)}          & \textbf{0.53}  & \textbf{135} \\
    \midrule
    \textbf{Grand total}           & \textbf{6.467} & \textbf{4{,}794.84} \\
    \bottomrule
  \end{tabular}
  \vspace{-4mm}
\end{table}

In aggregate, the logic occupies 5.94\,mm$^2$ (out of 6.47\,mm$^2$ total) and peaks at 4.66\,W; the aligner alone accounts for roughly three-quarters of both area and power, reflecting the single-pass similarity emphasis of the architecture. These synthesis numbers parameterize the cycle-accurate simulator used in the next section, where bank/precision gating and partial updates reduce average power to the 3.05–3.52\,W range while sustaining RT-60/RT-30 across all five tasks.

\subsection{Accelerator Execution Results}
All measurements come from the cycle-accurate simulator at 1\,GHz with switching activity taken from the 28\,nm synthesis. We begin with the latency envelope over the five tasks, then detail per-task runtime and energy, and finally compare throughput and energy with representative GPU pipelines on an RTX\,4090.

\subsubsection{Latency Envelope}
\begin{table}[h]
  \centering
  \small
  \caption{Min/max end-to-end per-frame latency across the five tasks (1\,GHz).}
  \vspace{-2mm}
  \label{tab:latency-envelope}
  \resizebox{\linewidth}{!}{%
  \begin{tabular}{lcccc}
    \toprule
    Mode & Global Min & Task (Min) & Global Max & Task (Max) \\
    \midrule
    RT-60 & 6.8\,ms  & have breakfast & 13.8\,ms & sports \\
    RT-30 & 12.9\,ms & have breakfast & 23.6\,ms & sports \\
    \bottomrule
  \end{tabular}
  }
  \vspace{-2mm}
\end{table}

The envelope shows comfortable headroom at both targets. Even the slowest frames remain below budget (13.8\,ms vs.\ 16.67\,ms at RT-60; 23.6\,ms vs.\ 33.33\,ms at RT-30), which is critical for absorbing DMA variance and host jitter. Tasks with higher temporal coherence such as \emph{have breakfast} consistently yield the lowest latencies, while motion-heavy scenes like \emph{sports} sit at the upper end of the range; this ordering matches the controller’s behavior of activating more banks and curtailing reuse when motion increases.

\subsubsection{Per-Task Runtime at RT Targets}
\begin{table}[h]
  \centering
  \small
  \caption{Accelerator runtime per task at RT targets. Latency in ms, power in W, energy in mJ per frame. Jitter = p95$-$median; Headroom = budget$-$p95 (16.67\,ms for RT-60; 33.33\,ms for RT-30).}
  \vspace{-2mm}
  \label{tab:accel-rt}
  \resizebox{\linewidth}{!}{%
  \begin{tabular}{lcccccccccccc}
    \toprule
    & \multicolumn{6}{c}{\textbf{RT-60 (60 FPS target)}} & \multicolumn{6}{c}{\textbf{RT-30 (30 FPS target)}} \\
    \cmidrule(r){2-7}\cmidrule(l){8-13}
    \textbf{Task} & Median & p95 & Jitter & Headroom & Power & Energy & Median & p95 & Jitter & Headroom & Power & Energy \\
    \midrule
    pour wine       & 9.4 & 11.3 & 1.9 & 5.37 & 3.20 & 53   & 17.2 & 19.9 & 2.7 & 13.43 & 3.50 & 116 \\
    sports          & 9.8 & 11.9 & 2.1 & 4.77 & 3.22 & 54   & 17.8 & 20.6 & 2.8 & 12.73 & 3.52 & 117 \\
    cooking         & 8.7 & 10.6 & 1.9 & 6.07 & 3.12 & 51   & 16.5 & 18.8 & 2.3 & 14.53 & 3.40 & 113 \\
    have breakfast  & 7.9 & 9.4  & 1.5 & 7.27 & 3.05 & 50   & 15.1 & 17.3 & 2.2 & 16.03 & 3.32 & 110 \\
    take a rest     & 8.1 & 9.7  & 1.6 & 6.97 & 3.06 & 50   & 15.4 & 17.6 & 2.2 & 15.73 & 3.33 & 110 \\
    \midrule
    \textbf{Average} & \textbf{8.78} & \textbf{10.58} & \textbf{1.80} & \textbf{6.09} & \textbf{3.13} & \textbf{51.6} & \textbf{16.40} & \textbf{18.84} & \textbf{2.44} & \textbf{14.49} & \textbf{3.41} & \textbf{113.2} \\
    \bottomrule
  \end{tabular}}
  \vspace{-2mm}
\end{table}

Across tasks, p95 latencies remain well within budget and jitter stays small (1.5–2.1\,ms at RT-60; 2.2–2.8\,ms at RT-30), indicating that the controller’s partial-update policy and $D'$ gating produce predictable service rather than bursty stalls. Energy per frame follows the expected scaling with the frame budget and the synthesis-calibrated power model: roughly 50–54\,mJ at RT-60 and 110–117\,mJ at RT-30. Scenes with more reuse (\emph{have breakfast}, \emph{take a rest}) exhibit the lowest median latency and energy, while high-motion scenes (\emph{sports}, \emph{pour wine}) require more active banks yet still leave milliseconds of headroom.

\subsubsection{Throughput/Power vs.\ GPU Baselines}
\begin{table}[h]
  \centering
  \small
  \caption{Throughput/energy (single RTX\,4090 vs.\ our accelerator) for representative task-oriented pipelines. GPU energy/frame is 450\,W divided by FPS.}
  \vspace{-2mm}
  \label{tab:gpu-vs-accel}
  \resizebox{\linewidth}{!}{%
  \begin{tabular}{lcccc}
    \toprule
    Method                 & Assumptions            & FPS (RTX\,4090) & Energy/Frame (RTX\,4090) \\
    \midrule
    TOIST (DETR)           & standard COCO input    & 15--25          & 30--18\,J \\
    iTaskCLIP (ViT-B/16)   & 120--200 crops/frame   & 5--12           & 90--38\,J \\
    iTaskCLIP (ViT-L/14)   & 120--200 crops/frame   & 2--6            & 225--75\,J \\
    \midrule
    \textbf{Ours (RT-60)}  & single-pass similarity & \textbf{60.3}   & \textbf{50\,mJ} \\
    \textbf{Ours (RT-30)}  & single-pass similarity & \textbf{30.1}   & \textbf{113\,mJ} \\
    \bottomrule
  \end{tabular}}
  \vspace{-2mm}
\end{table}

The comparison highlights why the accelerator carries the system end-to-end. Detector-only stacks on a 4090 land in the mid–teens to mid–twenties FPS; adding per-crop VLM alignment drives throughput into the low teens or single digits depending on backbone scale and crop count, with energy costs in the tens to hundreds of joules per frame. In contrast, the fixed-function design keeps all similarity work on dedicated datapaths, avoids re-encoding crops, and holds power nearly flat via $D'$ gating. The result is real-time throughput (30–60 FPS) at millijoule-scale energy, together with low jitter and consistent headroom across all five tasks.

\subsubsection{Execution Analysis}
\begin{itemize}
  \item \textbf{Frame-budget compliance:} p95 latency remains within \textbf{11--12\,ms} at RT-60 and \textbf{20--21\,ms} at RT-30 across all tasks (Table~\ref{tab:accel-rt}); the global envelope is 6.8--13.8\,ms (RT-60) and 12.9--23.6\,ms (RT-30) (Table~\ref{tab:latency-envelope}).
  \item \textbf{Energy advantage:} The accelerator sustains RT-60 at \textbf{$\sim$3.1\,W} average with \textbf{50\,mJ/frame}. Under comparable task pipelines, GPU baselines drop to single-digit FPS once per-crop VLM alignment (e.g., ViT-L/14) is included, with \textbf{56--225\,J/frame}. This gap underpins our system design.
  \item \textbf{Allocation strategy:} Steadier scenes (\emph{have breakfast}, \emph{take a rest}) allow more reuse and lower median latency; dynamic prompts (\emph{sports}, \emph{pour wine}) consume more compute yet remain comfortably within the frame budget.
\end{itemize}

\subsection{Task Accuracy vs. Prior Models}
\begin{table}[h]
  \centering
  \small
  \caption{AP@0.5 on the five evaluation tasks.}
  \vspace{-2mm}
  \label{tab:acc-sw}
  \resizebox{\linewidth}{!}{%
  \begin{tabular}{lccccc}
    \toprule
    Method & pour wine & sports & cooking & have breakfast & take a rest \\
    \midrule
    GGNN        & 40.7  & 43.6  & 37.6  & 39.1  & 40.5 \\
    TOIST       & 52.9  & 52.8  & 43.1  & 48.1  & 46.7 \\
    iTaskCLIP   & 63.51 & 65.54 & 56.08 & 44.39 & 44.76 \\
    iTaskCLIP*  & 63.36 & 62.02 & 56.14 & 42.59 & 45.62 \\
    \textbf{Ours (SW)} & \textbf{54.62} & \textbf{52.07} & \textbf{46.40} & \textbf{34.07} & \textbf{34.17} \\
    \bottomrule
  \end{tabular}
  }
  \vspace{-2mm}
\end{table}

Across these five tasks, our model is competitive where temporal coherence is higher and remains within a bounded margin to the strongest VLM baselines elsewhere. On the first two tasks, we achieve \textbf{54.62} and \textbf{52.07} AP—within \textbf{8.74} and \textbf{9.95} points of iTaskCLIP* and close to TOIST*. On the third task, we reach \textbf{46.40} AP, outperforming TOIST by \textbf{+3.3}. The latter two tasks are more challenging (\textbf{34.07}/\textbf{34.17}), yet the gaps to iTaskCLIP* remain moderate (\textbf{8.52}/\textbf{11.45}). Averaged over the five, our mean AP is \textbf{44.27\%} (vs.\ \textbf{53.95\%} for iTaskCLIP*), i.e., \textbf{75--86\%} of the strongest baseline per task while enabling the millijoule-scale, real-time execution demonstrated by the accelerator. This aligns with the system design: reuse-friendly scenes see the largest accuracy and energy benefits, and motion-heavy scenes keep tight p95 latency through controlled increases in active dimension without sacrificing frame-budget compliance.

\section{Conclusion}
We introduced \design{}, a cache-oriented algorithm–architecture co-design for task-oriented object detection that prioritizes \emph{compute-on-change}. By pairing a hyperdimensional associative reasoner with partial updates, dimension gating, and a lightweight real-time controller, the system sustains stable throughput on a tight power envelope, translating synthesis-calibrated cycle accuracy into predictable, deployment-ready behavior.

Looking forward, we see several paths to extend \design{}: finer-grained $D'$ and precision gating coupled with DVFS; learned runtime policies (bandits/RL) and differentiable scheduling; accuracy lift via task-aware distillation, quantization-aware training, and prompt-conditioned pruning; deeper memory work (bank placement, low-leakage SRAMs, lightweight compression, near-memory similarity); multi-prompt and multi-camera concurrency; migration to advanced nodes; and system-level validation with event cameras/IMUs, power-capped operation, and long-horizon continual adaptation. We plan to release reference RTL and evaluation traces to foster reproducibility and standardized benchmarking.

\bibliographystyle{IEEEtran}
\bibliography{IEEEabrv}

\end{document}